\documentclass[aps,prl,superscriptaddress,twocolumn,showpacs]{revtex4}
\usepackage{amsmath}
\usepackage{graphicx}
\usepackage{amssymb}
\usepackage{bbm, color, soul}
\usepackage{graphicx}
\usepackage{adjustbox}

\usepackage{color}

\usepackage{tikz}

\usepackage{amsthm}
\usepackage{tikz-cd}
\usepackage[all]{xy}
\usepackage{amsfonts}
\usepackage{mathrsfs}
\usepackage{hyperref}
\usepackage{MnSymbol} 

\theoremstyle{plain}

\theoremstyle{definition}
\newtheorem{defn/}[equation]{Definition}

\theoremstyle{remark}
\newtheorem{rmk/}[equation]{Remark}
\newtheorem*{rmk*}{Remark}

\newtheorem*{claim*}{Claim}

\begin{document}

\title[]{Local Wigner-Mass Maps and Integrated Negativity as Measures of nonclassicality in Quantum Chaotic Billiards}

\author{Kyu-Won \surname{Park}}
\email{parkkw7777@gmail.com}
\affiliation{Department of Mathematics and Research Institute for Basic Sciences, Kyung Hee University, Seoul, 02447, Korea}

\author{Soojoon \surname{Lee}}
\email{level@khu.ac.kr}
\affiliation{Department of Mathematics and Research Institute for Basic Sciences, Kyung Hee University, Seoul, 02447, Korea}
\affiliation{School of Computational Sciences, Korea Institute for Advanced Study, Seoul 02455, Korea}

\author{Kabgyun \surname{Jeong}}
\email{kgjeong6@snu.ac.kr}
\affiliation{Research Institute of Mathematics, Seoul National University, Seoul 08826, Korea}
\affiliation{School of Computational Sciences, Korea Institute for Advanced Study, Seoul 02455, Korea}

\date{\today}

\begin{abstract}
The Wigner function is a phase space quasi-probability distribution whose negative regions provide a direct, local signature of nonclassicality. To identify where phase-sensitive structure concentrates, we introduce local positive- and negative Wigner-mass maps and adopt the integrated Wigner negativity as a compact scalar measure of nonclassical phase space structure. A decomposition of the density operator reveals that off-diagonal coherences between hybridizing components generate oscillatory, sign-alternating patterns, with the negative contribution maximized when component weights are comparable. Non-Gaussian chaotic eigenmodes exhibit a baseline negativity that is further amplified by such hybridization. We validate these diagnostics across two billiard geometries and argue that the framework is transferable to other wave-chaotic platforms, where it can aid mode engineering and coherence control.
\end{abstract}

\maketitle

\section{Introduction}
The Wigner function provides a phase-space representation that reproduces the position and momentum marginals while encoding oscillatory cross terms that carry relative phase information \cite{Wigner1932,Moyal1949}. In contemporary experiments and theory, the Wigner distribution has been employed across diverse platforms, including optical Wigner tomography and displaced-parity measurements \cite{He2024,Kalash2023} and the motional states of trapped ions \cite{Leibfried1996,Fluehmann2020PRL}. It is also a convenient diagnostic in studies of quantum chaos \cite{Seidov2024,Cosic2020,Bandyopadhyay2021}. In particular, negative regions of the Wigner distribution provide a direct signature of nonclassicality \cite{Hillery1984,Kenfack2004}, and only Gaussian pure states are free of such negativity \cite{Hudson1974}. Wigner negativity has been observed and analysed in a variety of settings, including superconducting qubits \cite{Lu2021PRL} and resonators \cite{Hofheinz2009}, finite spin systems \cite{Davis2021}, and measurement-based witnesses that render the negativity operationally accessible \cite{Booth2022PRL,Chabaud2021}. Despite these advances, a quantitative understanding of how chaotic systems influence the generation and spatial distribution of Wigner negativity in realistic eigenmodes remains incomplete.

Quantum billiards provide a clean platform for studying interactions between eigenmodes, in particular avoided crossings (ACs). Small, controlled parameter changes bring near-degenerate eigenmodes into contact and reorganize nodal structure and phase-space weight \cite{Park2018,Park2022Chaos}. Such interactions are common in deformed cavities and produce families of partner modes with correlated features.

Conventional diagnostics based on eigenmode intensity \(|\psi(\mathbf r)|^{2}\) or on the Husimi \(Q\) function quantify mode spreading, and these measures are nonnegative. Shannon entropy computed from these measures can peak near an AC., reflecting increased spatial extent; yet these tools do not certify nonclassicality or isolate the growth of coherent superposition because phase information is suppressed \cite{Park2018,Park2022Chaos}. In our setting, non-Gaussian eigenmodes already carry a baseline negativity away from hybridization, and AC-driven off-diagonal coherence amplifies this intrinsic component.

Motivated by this gap, we develop a compact, spatially resolved framework based on Wigner negativity. Local maps of positive and negative Wigner mass pinpoint where phase-sensitive structure accumulates, and the integrated negativity provides a single scalar measure of the total nonclassical mass that is sensitive to eigenmode hybridization. A density-operator perspective links negativity to off-diagonal coherence between interacting components and explains why sign-changing structure grows when modal weights are comparable \cite{Cahill1969,Banaszek1999}. We apply these diagnostics to two billiard geometries and to parameter scans that induce hybridization \cite{Park2018,Park2022Chaos}. Mixed Wigner marginals retain a probabilistic interpretation, and reduced Wigner sections reveal sign-changing structure \cite{Wigner1932,Cohen1966}. The integrated negativity correlates quantitatively with a simple hybridization measure, and internal consistency is verified via marginal identities and the Wiener--Khinchin relation \cite{Hillery1984}. These maps and the scalar are readily evaluated from eigenmodes, and the framework is transferable to other wave-chaotic platforms where it may aid mode engineering and coherence control \cite{Besse2020PRX,He2024NatComm}.

The remainder of the paper is organized as follows. Section~II summarizes the Wigner–Weyl formalism. Section~III presents the Wigner–space analysis of eigenmodes. Section~IV shows representative maps, negativity traces, and correlation plots across ACs. Section~V interprets the results in terms of density-operator coherence. Section~VI concludes with a concise summary and outlook.

\section{Recapitulation of the Wigner function}

\noindent\textbf{Conventions.}
We work in two spatial dimensions and set $\hbar=1$ throughout.
We use the Fourier pair $\Psi(k)=\int d^2 r\, e^{-i\,k\cdot r}\,\psi(r)$ and $\psi(r)=\frac{1}{(2\pi)^2}\int d^2 k\, e^{+i\,k\cdot r}\,\Psi(k)$.
Parseval then gives $\int d^2 r\,|\psi(r)|^2=\frac{1}{(2\pi)^2}\int d^2 k\,|\Psi(k)|^2$.
All integrals use the Lebesgue measure on the indicated space, and all variables are real unless stated otherwise.

\subsection{Group-theoretic origin of the Wigner function}

We now recall the group-theoretic construction that underlies these conventions.
The Wigner-Weyl representation follows from the unitary irreducible representation of the Heisenberg-Weyl group \cite{Cahill1969,Hillery1984,Cohen1966}.
The canonical commutation relation is $[\hat q,\hat p]=i$.
A convenient displacement operator is $\hat D(q',p')=\exp\!\big[i\,(p'\hat q-q'\hat p)\big]$ \cite{Cahill1969}.
With the Hilbert-Schmidt inner product these operators form an orthogonal, complete operator frame \cite{Cahill1969,Hillery1984}.
We first state the orthogonality relation that makes the operator frame explicit.
In $d$ dimensions one has
\begin{align}
\mathrm{Tr}\!\big[\hat D^\dagger(q,p)\,\hat D(q',p')\big]
&=(2\pi)^{d}\,\delta(q-q')\,\delta(p-p') .
\end{align}
Any trace-class operator admits an expansion in this frame.
We then expand the density operator in the frame.
For the density operator we write
\begin{align}
\hat\rho
&=\iint \frac{dq'\,dp'}{(2\pi)^2}\,\chi(q',p')\,\hat D(q',p') ,
\\
\chi(q',p')
&=\mathrm{Tr}\!\big[\hat\rho\,\hat D^\dagger(q',p')\big] .
\end{align}
The Wigner function is the symplectic Fourier transform of the characteristic function.
With the sign convention that matches our Fourier pair, we define
\begin{align}
W(q,p)
&=\iint\frac{dq'\,dp'}{(2\pi)^2}\,\chi(q',p')\,e^{-i\,(p' q-q' p)} ,
\end{align}
which is consistent with the standard Cohen-class phase-space formalism \cite{Cohen1966}.
From these definitions one recovers reality for Hermitian $\hat\rho$, the correct normalization, and the correct marginals.

\subsection{Wigner distribution, marginals, and the role of Wiener–Khinchin}
\label{sec:phase-space}
\begin{figure*}
\centering
\includegraphics[width=15.5cm]{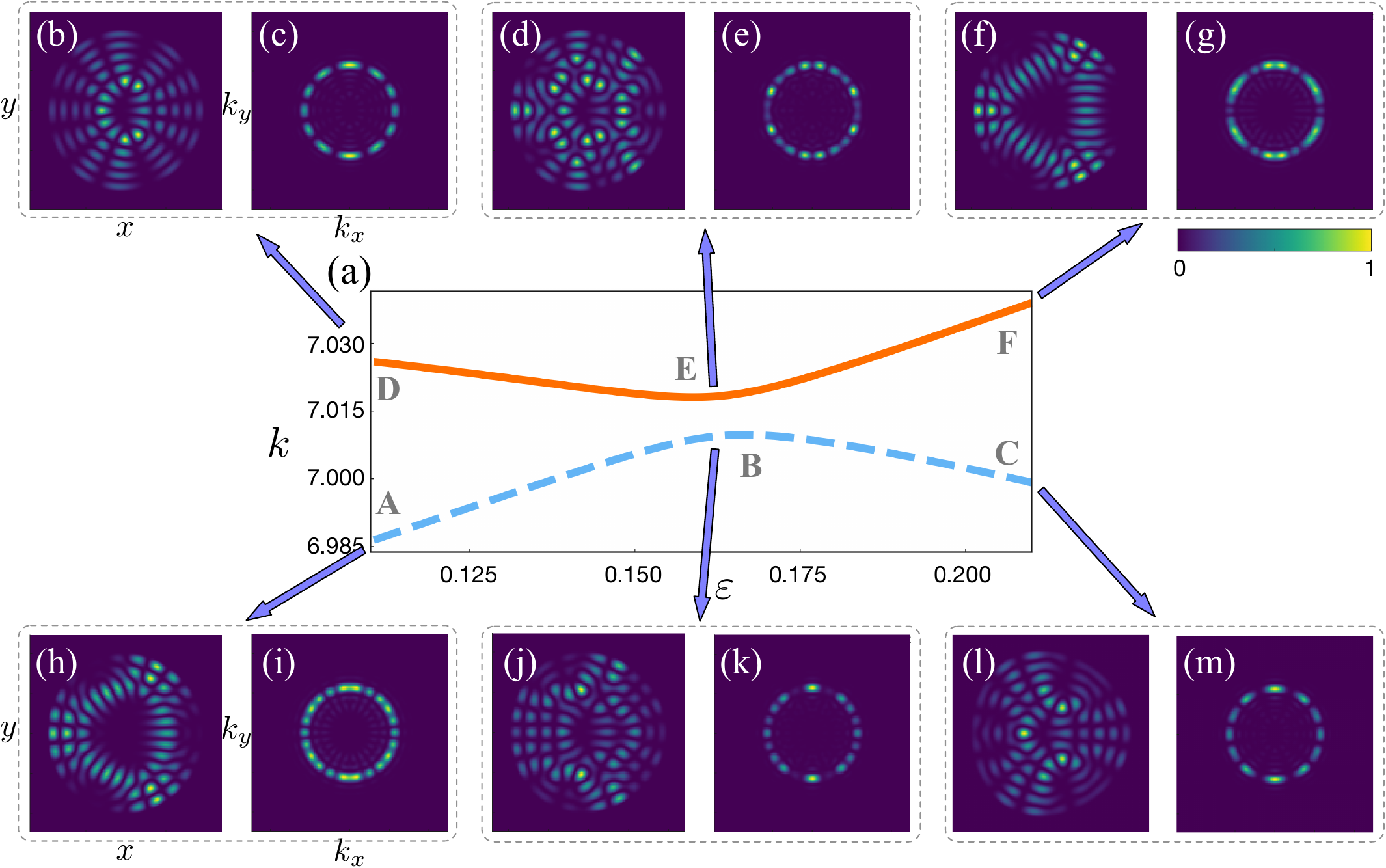}
\caption{Spectral trajectories and representative eigenmodes through an avoided crossing. (a) Eigenvalue trajectories $k$ versus deformation $\varepsilon$; markers A--F indicate sampled parameter values. The solid orange curve in (a) corresponds to mode~2, and the dashed cyan curve corresponds to mode~1. (b)--(m) Paired position $|\psi(r)|^2$ (left or upper) and momentum-space $(2\pi)^{-2}|\Psi(k)|^2$ (right or lower) intensity maps. Near the AC.\ center, the position intensity exhibits increased mixing and delocalization, while momentum-space signatures retain the characteristic ring/lobed structure of each eigenmode; the momentum-ring radius remains nearly constant at $\approx 22$.}
\label{Figure-1}
\end{figure*}

We now pass to center and relative coordinates.
Let $R=(r+r')/2$ and let $q'=r'-r$.
The position representation of the matrix element of $\hat D^\dagger$ is
\begin{align}
\langle r'|\hat D^\dagger(q',p')|r\rangle
&= e^{-i\,p'(r+r')/2}\,\delta(r'-r-q') .
\end{align}
We insert this expression into the center–relative form of the characteristic function, and then substitute the result into the symplectic Fourier transform for $W$.
The $p'$ integral yields
\begin{align}
\int d^2 p'\,e^{\,i\,p'\cdot(r-R)}=(2\pi)^{2}\,\delta^{(2)}(r-R)
\end{align}
The center integral collapses, and we relabel $q'\mapsto s$ and $p'\mapsto k$.
For a pure state with $\hat\rho=|\psi\rangle\langle\psi|$ we obtain the relative-coordinate representation used in the numerical sections,
\begin{align}
W(r,k)
&=\frac{1}{(2\pi)^2}\int d^2 s\,
\psi\!\left(r+\tfrac{s}{2}\right)\psi^*\!\left(r-\tfrac{s}{2}\right)\,
e^{-i\,k\cdot s} .
\label{eq:W_def}
\end{align}
The marginals follow directly from this form: integrating over $k$ yields $\int d^2k,W(r,k)=|\psi(r)|^2$, while integrating over $r$ yields $\int d^2r,W(r,k)=\tfrac{1}{(2\pi)^2}|\Psi(k)|^2$.
If $\int d^2 r\,|\psi|^2=1$ then Parseval implies $(2\pi)^{-2}\int d^2 k\,|\Psi|^2=1$.
Hence
\begin{align}
\iint d^2 r\,d^2 k\,W(r,k)&=1 .
\end{align}

We next state the Wiener-Khinchin pair that we use as an internal consistency check \cite{Wiener1930,Khinchin1934}.
In our numerical pipeline the Wiener–Khinchin relation serves as a strict internal check and is not used as a definition of $\Psi$.
We define the configuration-space autocorrelation
\begin{align}
R(s)
&=\int d^2 r\,
\psi\!\left(r+\tfrac{s}{2}\right)\psi^*\!\left(r-\tfrac{s}{2}\right) .
\end{align}
With the conventions stated above the Fourier pair holds with the kernel $e^{-i\,k\cdot s}$,
\begin{align}
\frac{1}{(2\pi)^2}\,|\Psi(k)|^2
\quad\longleftrightarrow\quad
R(s) .
\label{eq:WK}
\end{align}
Verifying this identity together with the two marginal identities confirms that FFT normalizations, grid spacings, and integration measures are mutually consistent to machine precision. We now illustrate these definitions and checks on the oval-billiard data shown in Fig.~\ref{Figure-1}.

\section{Wigner-Space Analysis of Eigenmodes in quantum chaotic billiards}
This section examines eigenmode structure through Wigner phase-space representations, emphasizing modal hybridization and momentum signatures in chaotic billiards.

Classically, a billiard is a point particle undergoing free motion interrupted by specular reflections at a hard-wall boundary; the dynamics are generated by the Hamiltonian $\mathcal{H}(q,p)=p^2/2m+V(q)$ with $V=0$ inside the domain and $V=\infty$ on the wall. The resulting Hamiltonian flow conserves phase-space volume and, for the geometries studied here, is strongly nonintegrable and mixing, providing the organizing structure for semiclassical intuition \cite{Sinai1970,Bunimovich1979,Berry1977}.

Quantum mechanically, we solve the time-independent Schr\"odinger (Helmholtz) problem under Dirichlet boundary conditions. In the closed-cavity (quantum billiard) the governing equation is
\begin{align}
(\nabla^2 + n^{2}k^2)\,\psi(\mathbf r)=0,
\end{align}
where $k$ is the vacuum wavenumber and the physical spatial frequency inside the resonator is $n k$ (the internal wavenumber scales linearly with the refractive index $n$). In the calculations reported below we assume TM polarization and $n=3.3$; $\psi$ denotes the $z$-component of the electric field.

First, we consider the oval-shaped chaotic quantum billiard obtained by deforming an ellipse in the $x$-direction,
\begin{align}
\frac{x^2}{a^2}+\bigl(1+\varepsilon x\bigr)\frac{y^2}{b^2}=1,
\end{align}
with $a=1.0,\; b=1.03$. The deformation breaks integrability and produces the level repulsion and complex modal structure characteristic of quantum-chaotic systems \cite{Berry1977,Park2025}.

\begin{figure*}[htbp]
\centering
\includegraphics[width=15.5cm]{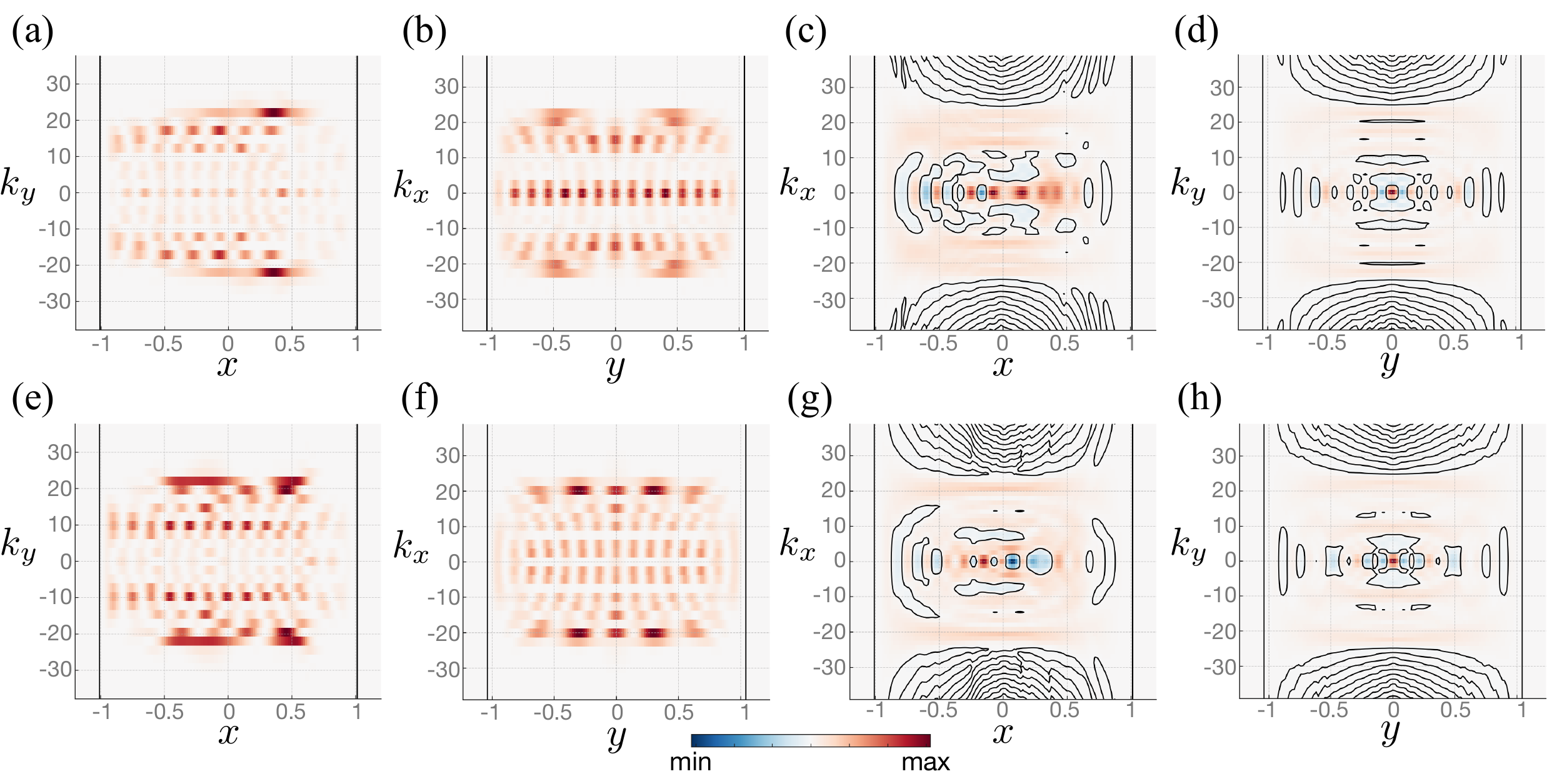}
\caption{Wigner marginals for the two interacting modes near the avoided crossing. Panels (a)--(d) show Mode E in Figure~\ref{Figure-1} and panels (e)--(h) show Mode B in Figure~\ref{Figure-1} . For each mode the four marginal projections are presented in the same order: (a),(e) $f(x,k_y)$; (b),(f) $f(y,k_x)$; (c),(g) $f(x,k_x)=W(x,k_x)$ (reduced Wigner); (d),(h) $f(y,k_y)=W(y,k_y)$ (reduced Wigner). Accordingly, the mixed marginals $f(x,k_y)$ and $f(y,k_x)$ are always non-negative and reproduce Born-rule probability densities, whereas the reduced Wigner projections $f(x,k_x)$ and $f(y,k_y)$ may exhibit negative lobes (blue regions) that signal quantum interference and nonclassical phase-space structure. Black contour lines indicate the zero level of the reduced Wigner functions ($f=0$) and serve as a visual guide. A common colorbar applies to all panels.}
\label{Figure-2}
\end{figure*}

Eigenvalues and eigenfunctions are computed with the boundary element method (BEM) under Dirichlet conditions \cite{Wiersig02}. By varying the deformation parameter $\varepsilon$, we trace spectral trajectories, sample representative modes, and identify avoided crossings that are analyzed in phase space via Wigner distributions and negativity measures.

Figure~1 summarizes the spectral evolution and representative eigenmodes through a single avoided crossing (AC). The central panel (Fig.~1a) shows the two interacting eigenvalue branches plotted as functions of the deformation parameter $\varepsilon$; labeled markers A--F indicate the parameter values at which modes were sampled. In Fig.~1a the solid orange curve corresponds to mode~2, while the dashed cyan curve corresponds to mode~1. Panels (b)--(m) present paired configuration- and momentum-space intensities for those samples: in each pair the left panel displays the normalized configuration intensity $\int W(r,k)\,dk_x\,dk_y = |\psi(r)|^2$, and the right panel shows the corresponding momentum (spatial-frequency) intensity $\int W(r,k)\,dx\,dy = \frac{1}{(2\pi)^2}|\Psi(k)|^2$. Far from the interaction each branch exhibits a stable nodal pattern and a characteristic momentum-space ring or lobed structure, consistent with a single-character eigenmode of definite parity. Approaching the AC center, the configuration intensity clearly hybridizes: nodal lines rearrange and $|\psi|^2$ becomes more mixed and spatially extended (increased delocalization), whereas the momentum-space signature of each eigenmode preserves its ring/lobed form because the modal spatial frequency is fixed by the eigenvalue.

In addition, the theoretical internal wavenumber is given by \(n k\) for the chosen refractive index and eigenvalue. The relative deviation between the measured peak and the theoretical internal wavenumber is small, indicating excellent agreement between the measured momentum-space peak and the expected value.

\begin{figure*}
\centering
\includegraphics[width=16.5cm]{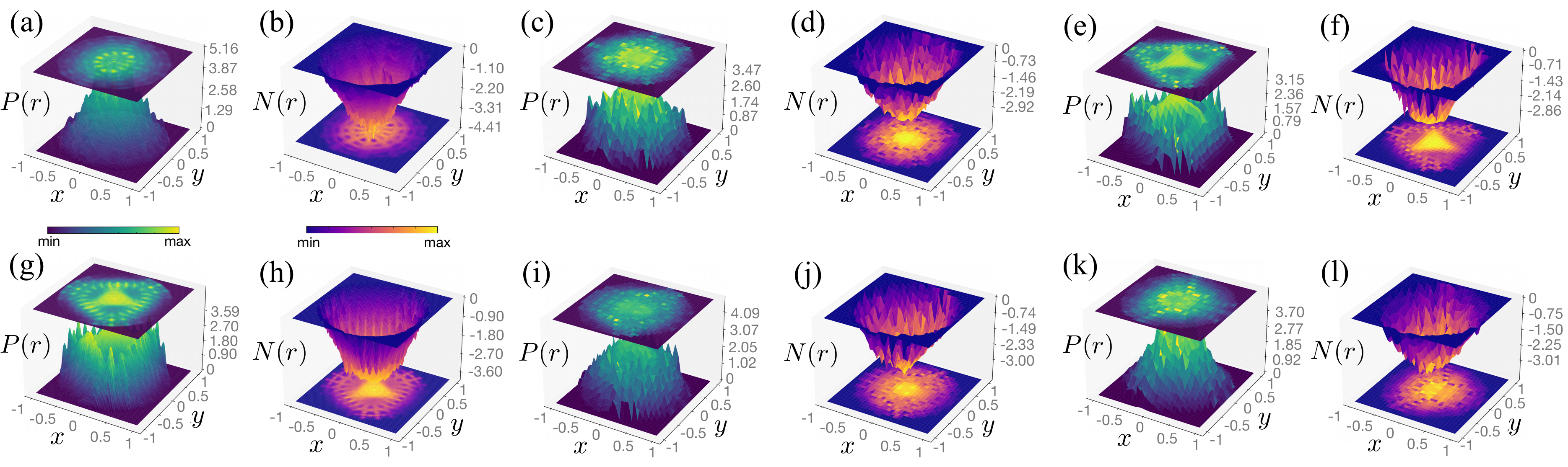}
\caption{Local positive and negative phase space masses, \(P(r)\) (green) and \(N(r)\) (orange), shown as 3D surfaces over the configuration space \((x,y)\). Panels (a,b), (c,d), and (e,f) correspond to eigenmodes D, E and F from Fig.~1(a), while (g,h), (i,j) and (k,l) correspond to A, B and C. In each pair the left surface is \(P(r)\) and the right is \(N(r)\). Across all parameter points \(P(r)\) is slightly larger in magnitude than \(N(r)\), consistent with the global Wigner negativity. The spatial maps display magnitudes that tend to increase toward the configuration space origin \((r=0)\), indicating stronger local phase space mass near the center. These maps provide a spatial decomposition of the global nonclassicality measure \(\mathcal{N}\).}
\label{Figure-3}
\end{figure*}

\section{Mixed marginals and reduced Wigner sections}
\label{sec:fig2-mixed-vs-reduced}
Building on Sec.~\ref{sec:phase-space}, we now contrast mixed marginals with reduced Wigner sections.
Figure~\ref{Figure-2} compares the mixed Wigner marginals \(f(x,k_y),\,f(y,k_x)\) with the reduced
Wigner sections \(f(x,k_x),\,f(y,k_y)\) for the two interacting modes shown in Fig.~\ref{Figure-1}.
For compactness we project the full Wigner function \(W(x,y,k_x,k_y)\) as follows:
\begin{align}
f(x,k_y)=\int W\,dk_x\,dy,\qquad f(y,k_x)=\int W\,dk_y\,dx,
\end{align}
\begin{align}
f(x,k_x)=\int W\,dy\,dk_y,\qquad f(y,k_y)=\int W\,dx\,dk_x.
\end{align}
The first pair \((f(x,k_y),f(y,k_x))\) are \emph{mixed} marginals involving one real-space
and one orthogonal momentum coordinate. The second pair \((f(x,k_x),f(y,k_y))\) are
\emph{reduced} Wigner sections on the canonical planes \((x,k_x)\) and \((y,k_y)\).

Each mixed marginal is a one-dimensional power spectrum obtained from an autocorrelation
and is therefore nonnegative by the Wiener–Khinchin theorem~\cite{Wiener1930,Khinchin1934}.
Define the short autocorrelations and their Fourier transforms; the following compact relations then hold.
\begin{align}
R_y(x;s) &:= \int \psi\!\bigl(x,y+\tfrac{s}{2}\bigr)\,
             \psi^*\!\bigl(x,y-\tfrac{s}{2}\bigr)\,dy,
\label{eq:Ry} \\[4pt]
R_x(y;s) &:= \int \psi\!\bigl(x+\tfrac{s}{2},y\bigr)\,
             \psi^*\!\bigl(x-\tfrac{s}{2},y\bigr)\,dx.
\label{eq:Rx}
\end{align}
\begin{align}
f(x,k_y) &= \dfrac{1}{2\pi}\,\widehat{R}_y(x,k_y)
          \;=\; \dfrac{1}{2\pi}\,\bigl|\mathcal{F}_y\{\psi(x,\cdot)\}(k_y)\bigr|^2
          \;\ge\; 0,
\label{eq:fxpy} \\[6pt]
f(y,k_x) &= \dfrac{1}{2\pi}\,\widehat{R}_x(y,k_x)
          \;=\; \dfrac{1}{2\pi}\,\bigl|\mathcal{F}_x\{\psi(\cdot,y)\}(k_x)\bigr|^2
          \;\ge\; 0.
\label{eq:fypx}
\end{align}
Here \(\widehat{R}_y\) and \(\widehat{R}_x\) denote the Fourier transforms of \(R_y\) and
\(R_x\) with respect to the lag variable \(s\). Equations \eqref{eq:fxpy}--\eqref{eq:fypx}
are the Wiener–Khinchin statements written compactly and demonstrate that the mixed
marginals are nonnegative.

By contrast, the reduced projections
\begin{align}
f(x,k_x)=\int W\,dy\,dk_y,\qquad f(y,k_y)=\int W\,dx\,dk_x
\end{align}
are genuine Wigner sections on conjugate planes. The Wigner kernel is bilinear in
\(\psi\) and contains cross (interference) terms between spatially separated components.
Consequently these reduced distributions may assume negative values. In
Fig.~\ref{Figure-2}(c),(g) and (d),(h) the negative lobes (blue) are bounded by the \(f=0\) contours,
whereas the mixed panels in Fig.~\ref{Figure-2}(a),(e) and (b),(f) remain strictly nonnegative as
guaranteed by the Wiener–Khinchin relation.

Taken together, these plots show that across the avoided crossing the mixed marginals
remain nonnegative and thus retain their interpretation as probability densities, while the reduced Wigner sections expose parameter-dependent, sign-changing interference patterns that reflect the nonclassical phase-space structure of the hybridized modes.

Moreover, the sign-changing structure observed in the reduced Wigner sections is neither spatially nor spectrally uniform. The negative lobes in the reduced projections concentrate about the origin in momentum space \((k_x,k_y\approx 0)\) and are spatially localized near the configuration-space origin \((x,y\approx 0)\). In other words, the principal contribution to the reduced Wigner negativity arises from origin-centered momentum features that lie close to the cavity centre in real space. We examine this origin-centred localization quantitatively in the next section (Sec.~V), where configuration-space negative mass maps \(N(r)\) and momentum-space negative mass maps \(\widetilde N(k)\) make the association between low-\(k\) structure and the global negativity \(\mathcal{N}\) explicit.

\begin{figure*}
\centering
\includegraphics[width=16.5cm]{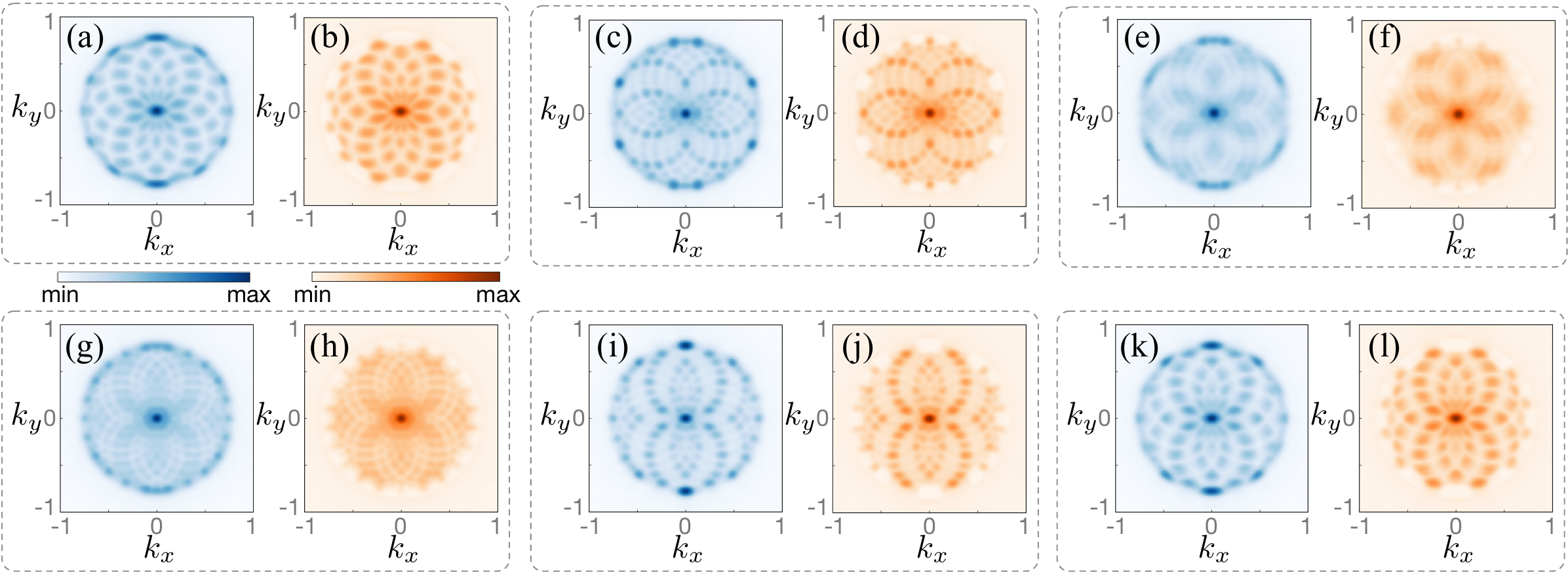}
\caption{Momentum-space local positive and negative phase-space masses, \(\widetilde P(k)\) (blue) and \(\widetilde N(k)\) (orange), shown as intensity maps over momentum coordinates \((k_x,k_y)\). Panels (a,b), (c,d), and (e,f) correspond to eigenmodes D, E and F from Fig.~\ref{Figure-1}(a), while (g,h), (i,j) and (k,l) correspond to A, B and C. In each pair the left image is \(\widetilde P(k)\) and the right image is \(\widetilde N(k)\). Although the integrated positive contribution slightly exceeds the negative one as in Fig.~\ref{Figure-3}, the momentum-space maps exhibit a pronounced negative peak at the origin \((k=0)\), indicating that the dominant negative phase-space mass is concentrated at zero momentum. These plots complement the configuration-space decomposition of the global Wigner negativity \(\mathcal{N}\).}
\label{Figure-4}
\end{figure*}

\section{Wigner negativity definition and structure}
\label{sec:negativity}
With the Wigner formalism established, we now quantify and localize nonclassicality through the negativity and its spatial and spectral decompositions.

\subsection{Global definition and local mass decomposition}

The phase-space nonclassicality of an eigenmode is quantified by the Wigner negativity~\cite{Kenfack2004}. We adopt the sign-explicit form
\begin{equation}
\mathcal{N} \;=\; \tfrac{1}{2}\iint\bigl(|W(r,k)|-W(r,k)\bigr)\,d^2r\,d^2k,
\label{eq:neg-def}
\end{equation}
so that \(\mathcal{N}\) integrates the negative lobes of \(W\) and vanishes iff \(W\ge 0\) almost everywhere. Using \(\iint W\,d^2r\,d^2k=1\), this is algebraically equivalent to \(\mathcal{N}=\tfrac{1}{2}\bigl(\iint |W|\,d^2r\,d^2k-1\bigr)\).

Because \(\mathcal{N}\) is sensitive to oscillatory interference, it provides an efficient probe of eigenmode hybridization near AC\ points: hybridized modes generate stronger phase-space oscillations and hence larger negative mass.

To expose the spatial structure underlying the global diagnostic $\mathcal{N}$, we split the Wigner function into its positive and negative parts, $W_{+}(r,k)=\max\{W(r,k),0\}$ and $W_{-}(r,k)=-\min\{W(r,k),0\}$, and define the corresponding local (configuration-space) masses by
\begin{align}
P(r)&:=\int W_{+}(r,k)\,\mathrm{d}^{2}k,\label{eq:P_def}\\
N(r)&:=\int W_{-}(r,k)\,\mathrm{d}^{2}k.\label{eq:N_def}
\end{align}
Here $P(r)$ and $N(r)$ give, respectively, the positive and negative phase-space mass concentrated at position $r$.

By construction,
\begin{align}
P(r)-N(r)=\int W(r,k)\,d^2k = |\psi(r)|^2 \qquad\text{for every } r,
\end{align}
where \(\psi(r)\) denotes the eigenmode amplitude in configuration space. Integrating over configuration space reproduces the normalization condition.

We now examine how the integrated positive and negative masses vary with the oval deformation parameter \(\epsilon\).
\begin{figure}
\centering
\includegraphics[width=7.0cm]{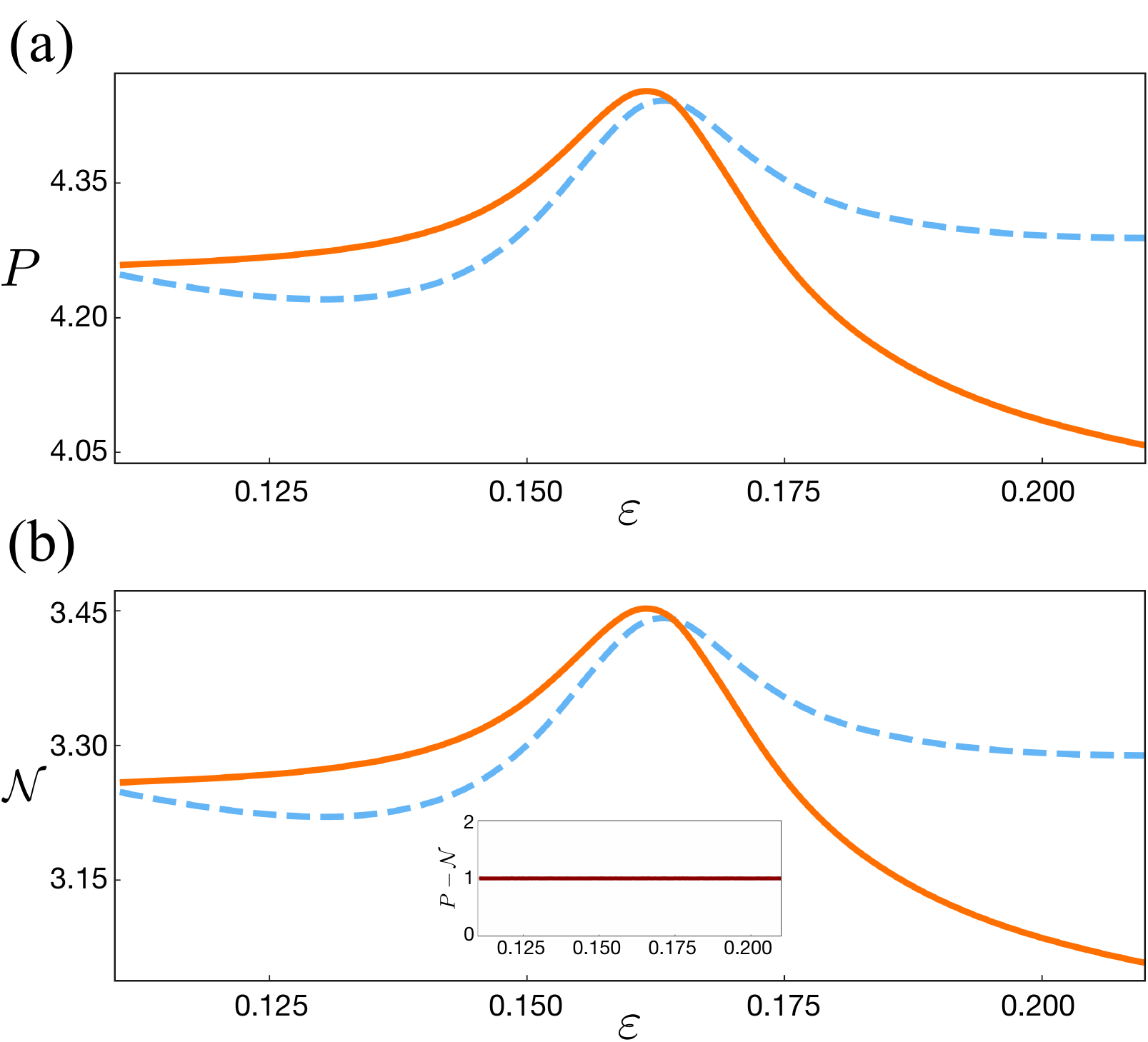}
\caption{%
(a)~Integrated positive mass \(P=\int P(r)\,d^2r\) and (b)~negative mass \(\mathcal{N}=\int N(r)\,d^2r\) as functions of the deformation parameter \(\epsilon\) for the avoided-crossing pair in Fig.~1(a).
Both quantities peak near the AC., reflecting enhanced phase-space interference.
The inset in (b) verifies the normalization \(P-N=1\) across the full range of~\(\epsilon\).}
\label{Figure-5}
\end{figure}
Crucially, the global negativity equals the spatial integral of the local negative mass as follows:
\begin{equation}
\mathcal{N} \;=\; \int N(r)\,d^2r.
\label{eq:N_local_global}
\end{equation}
Thus maps of \(N(r)\) (and complementary maps of \(P(r)\)) provide a direct, local decomposition of the global quantity \(\mathcal{N}\): spatial regions with large \(N(r)\) are precisely those where the eigenmode departs most strongly from any classical, nonnegative phase-space description.

Figure~\ref{Figure-3} presents these configuration-space maps as 3D surfaces: \(P(r)\) (green) and \(N(r)\) (orange), shown pairwise for the representative eigenmodes. Across the parameter points displayed, the local positive mass \(P(r)\) typically slightly exceeds the local negative mass \(N(r)\), consistent with the global balance $\int P(r)\,d^2r-\int N(r)\,d^2r=1$. The maps also show that both \(P(r)\) and \(N(r)\) often grow toward the configuration-space origin \(r=0\), indicating stronger phase-space mass near the cavity center. Regions where \(N(r)\) is large coincide with strongly oscillatory Wigner slices. Conversely, where \(N(r)\) is small the positive mass \(P(r)\) better traces the coarse features of the eigenmode intensity \(|\psi(r)|^2\).

\subsection{Spatial and momentum localization in the oval billiard}

Complementing the spatial decomposition, Fig.~\ref{Figure-4} shows the momentum-space analogues \(\widetilde P(k)\) and \(\widetilde N(k)\), obtained by integrating \(W_+\) and \(W_-\) over configuration space. In all examined cases, the momentum-space negative mass \(\widetilde N(k)\) displays a pronounced concentration at the momentum origin \(k=0\): a strong negative peak at low \(k\) that supplies the dominant contribution to \(\mathcal{N}\). Because \(\mathcal{N}=\int\widetilde N(k)\,d^2k\), these origin-centered features identify the principal \(k\)-region responsible for nonclassicality and clarify how changes in \(\mathcal{N}\) are tied to specific momentum-space structures of interacting eigenmodes. Together, the \(P(r),N(r)\) and \(\widetilde P(k),\widetilde N(k)\) maps give a compact, two-space localization of the global Wigner negativity and make transparent the spatial and momentum fingerprints of hybridization near AC\ points.

Figure~\ref{Figure-5} presents the global counterparts of the local quantities shown in Figs.~\ref{Figure-3}--\ref{Figure-4}.
Here the integrated masses
\begin{align}
P(\epsilon)=\int P(r;\epsilon)\,d^2r,\qquad \mathcal{N}(\epsilon)=\int N(r;\epsilon)\,d^2r
\end{align}
quantify how the total positive and negative phase-space weights evolve with the deformation parameter \(\epsilon\).
Both curves exhibit a pronounced maximum near the avoided crossing, consistent with the expectation that eigenmode hybridization amplifies oscillatory structure in \(W(r,k)\) and therefore increases the overall Wigner negativity.

\subsection{Interference mechanism and geometry dependence}
\begin{figure*}
\centering
\includegraphics[width=16.5cm]{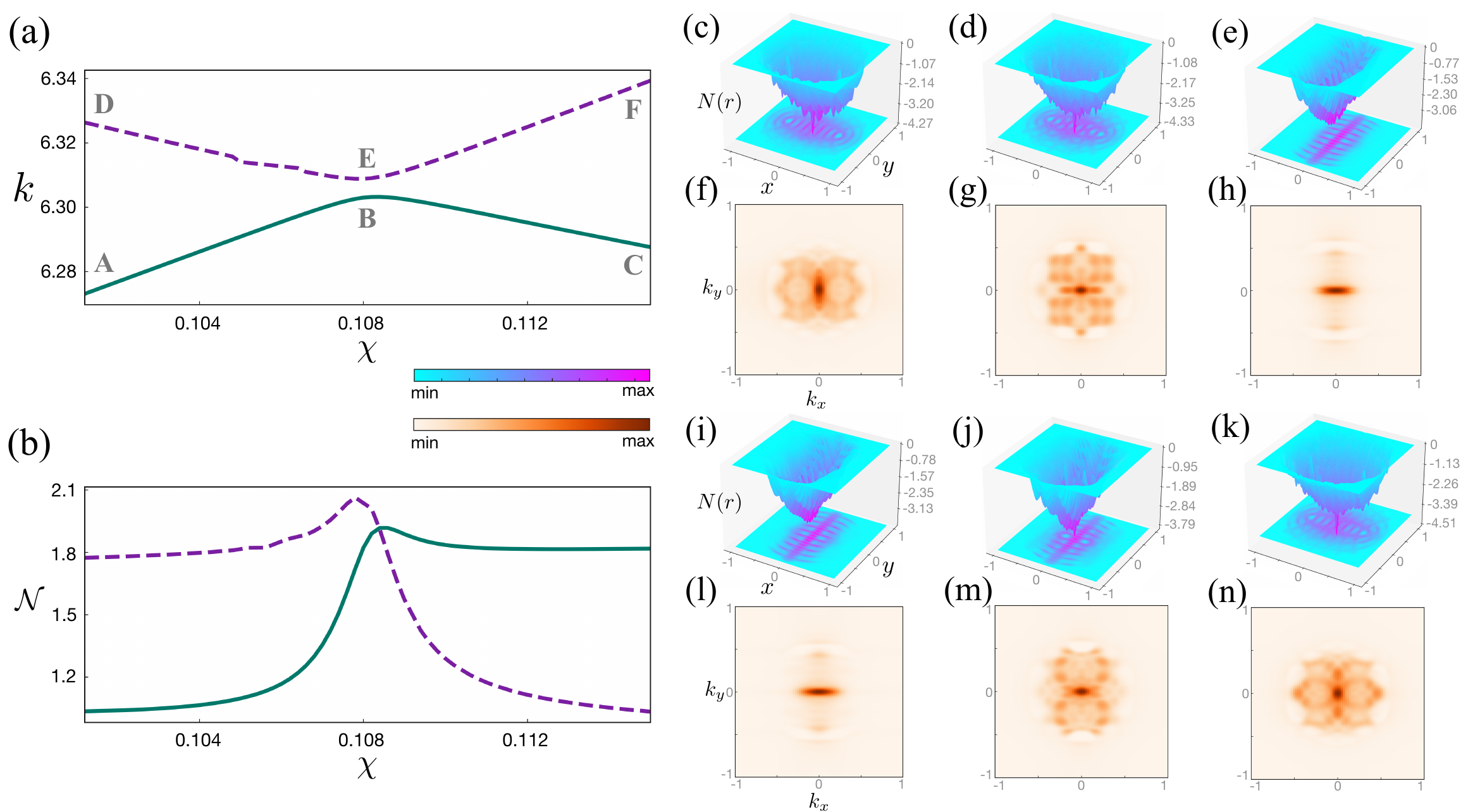}
\caption[Quadrupole billiard maps]{Quadrupole-billiard results. (a) Eigenvalue trajectories $k(\chi)$ showing the interacting A--C pair; solid and dashed curves mark the two branches involved in the interaction. (b) Total Wigner negativity $\mathcal{N}$ as a function of the deformation parameter $\chi$. Panels (c--e) and (i--k) present configuration-space maps $N(r)$ as three-dimensional surfaces for eigenmodes D, E, F and A, B, C, respectively, while the corresponding momentum-space maps $N(k)$ are shown below in (f--h) and (l--n) as two-dimensional intensity plots. Panel correspondence is: (c,f)$\rightarrow$D, (d,g)$\rightarrow$E, (e,h)$\rightarrow$F, (i,l)$\rightarrow$A, (j,m)$\rightarrow$B, and (k,n)$\rightarrow$C. These quadrupole-billiard maps display patterns qualitatively similar to those observed for the oval billiard; Figs.\ 4--5 indicate comparable spatial and momentum-space structure of the local negative phase-space mass.}
\label{Figure-6}
\end{figure*}

This behaviour admits a transparent description in terms of the density operator and its decomposition into diagonal and off-diagonal (coherence) sectors.
Writing the pure-state density matrix in the position basis and setting \(\psi=\alpha\psi_1+\beta\psi_2\) yields
\begin{equation}
\rho= |\alpha|^2\rho_{11}+|\beta|^2\rho_{22} + \alpha\beta^*\,\rho_{12} + \alpha^*\beta\,\rho_{21},
\end{equation}
with \(\rho_{ij}(r,r')=\psi_i(r)\psi_j^*(r')\). The first two (diagonal) blocks encode populations, while the last two (off-diagonal) blocks carry coherence responsible for interference in phase space~\cite{NielsenChuang2000}.

The Wigner transform acts on \(\rho\) by Fourier transforming the relative coordinate \(m=(r-r')/2\), so each matrix block \(\rho_{ij}\) yields a corresponding phase-space component
\begin{align}
W_{ij}(r,k)=\int \psi_i(r+m)\,\psi_j^*(r-m)\,e^{-2ik\cdot m}\,d^2m,
\end{align}
and therefore
\begin{widetext}
\begin{equation}
W(r,k)=|\alpha|^2 W_{11}(r,k)+|\beta|^2 W_{22}(r,k)
+\alpha\beta^* W_{12}(r,k)+\alpha^*\beta W_{21}(r,k).
\end{equation}
\end{widetext}
Taking the real part recovers the familiar form
\begin{align}
W=|\alpha|^2W_1+|\beta|^2W_2+2\Re\{\alpha\beta^*W_{12}\},
\end{align}
where the off-diagonal transforms \(W_{12},W_{21}=W_{12}^*\) encode the inter-component coherence. Because \(W_{12}\) is the Fourier transform of the off-diagonal density \(\rho_{12}(r+m,r-m)\), it is generically oscillatory in \(r\) and \(k\); hence the interference term \(2\Re\{\alpha\beta^*W_{12}\}\) produces alternating positive/negative lobes in phase space. Its amplitude scales as \(2|\alpha\beta||W_{12}|\), and under the normalization \(|\alpha|^2+|\beta|^2=1\) the prefactor \(|\alpha\beta|\) is maximized at \(|\alpha|=|\beta|=1/\sqrt{2}\) (giving \(|\alpha\beta|_{\max}=1/2\)) and vanishes in the limit \(|\alpha|\to1,|\beta|\to0\) (or vice versa). Thus the density-operator viewpoint makes transparent why off-diagonal coherence terms are the phase-space carriers of interference and why hybridization (comparable component weights) maximally amplifies the cross-term that generates Wigner negativity.

Second, appreciable negativity away from the AC is expected on independent grounds. By Hudson's theorem a pure state has a nonnegative Wigner function if and only if it is Gaussian\cite{Hudson1974}. Therefore any intrinsically non-Gaussian eigenmode carries a baseline Wigner negativity even in the absence of mode mixing. In practice the observed \(N(\epsilon)\) therefore combines an intrinsic part (due to non-Gaussian structure of individual modes) and an interference part (due to cross-terms amplified by hybridization near the AC), with the pronounced peak in Fig.~\ref{Figure-5} corresponding primarily to the latter superposed on the former baseline. Finally, the inset in Fig.~\ref{Figure-5}(b) verifies the balance \(P(\epsilon)-\mathcal{N}(\epsilon)=1\) across the parameter range, confirming the normalization condition.

To assess geometry dependence we repeat the same analysis for the quadrupole billiard parameterized by \(\chi\).
Figure~\ref{Figure-6} shows results for the quadrupole billiard whose boundary is defined in polar coordinates as
\begin{align}
\rho(\theta) = 1 + \chi \cos(2\theta).
\end{align}
Panel (a) shows eigenvalue trajectories \(k(\chi)\) highlighting the interacting A--C pair, and (b) gives the total Wigner negativity \(\mathcal{N}(\chi)\). Panels (c--e) and (i--k) display configuration-space maps \(N(r)\) for modes D--F and A--C, respectively, while (f--h) and (l--n) show the corresponding momentum-space maps \(N(k)\); the explicit pairing is (c,f)\(\to\)D, (d,g)\(\to\)E, (e,h)\(\to\)F, (i,l)\(\to\)A, (j,m)\(\to\)B, (k,n)\(\to\)C. The quadrupole maps reproduce the same localization motifs and momentum-space concentration of negative phase-space mass previously observed for the oval billiard, demonstrating that these features persist across different boundary deformations. This agreement indicates the robustness of negative-phase-space formation under generic mode hybridization, suggesting the effect is a universal consequence of the interaction dynamics rather than a geometry-specific artifact.

\section*{Conclusion}
We introduced a compact framework that employs Wigner negativity to probe phase space structure in chaotic quantum billiards. Local positive and negative Wigner-mass maps locate phase-sensitive features, while the integrated negativity provides a single scalar quantifying hybridization strength. Using a density-operator decomposition, we showed that off-diagonal coherence between hybridizing components produces oscillatory, sign-alternating contributions that are amplified when component weights are comparable.

By Hudson's theorem, only Gaussian pure states are everywhere nonnegative; hence intrinsically non-Gaussian eigenmodes carry a baseline negativity, and hybridization adds an interference term that produces pronounced peaks at avoided crossings. The local negative mass is localized in phase space: momentum-space negativity concentrates near the momentum origin (low-\(k\)), while configuration-space negativity grows toward the cavity center.

These results demonstrate that Wigner negativity is a practical, numerically testable diagnostic for phase-sensitive structure in chaotic billiards and our result requires only the Wigner function (computable from measured or simulated field data), the proposed diagnostics readily extend to optical, acoustic, and microwave resonators, and can guide targeted mode shaping in practical devices.”

\section{acknowledgement}
This work was supported by the National Research Foundation of Korea (NRF) through a grant funded by the Ministry of Science and ICT (Grants Nos. RS-2023-00211817 and RS-2025-00515537), the Institute for Information \& Communications Technology Promotion (IITP) grant funded by the Korean government (MSIP) (Grants No. RS-2025-02304540), and the National Research Council of Science \& Technology (NST) (Grant No. GTL25011-401). S.L. acknowledges support from the National Research Foundation of Korea (NRF) grants funded by the MSIT (Grant No. RS-2022-NR068791).


\end{document}